# LOCALITY SIM: CLOUD SIMULATOR WITH DATA LOCALITY

Ahmed H.Abase[1], Mohamed H. Khafagy[2] and Fatma A. Omara[3]

[1]Computer Science Department, Cairo University, EGYPT
[2]Computer Science Department, Fayoum University, EGYPT
[3]Computer Science Department, Cairo University, EGYPT

## ABSTRACT

*Cloud Computing (CC) is a model for enabling on-demand access to a shared pool of configurable computing resources. Testing and evaluating the performance of the cloud environment for allocating, provisioning, scheduling, and data allocation policy have great attention to be achieved. Therefore, using cloud simulator would save time and money, and provide a flexible environment to evaluate new research work. Unfortunately, the current simulators (e.g., CloudSim, NetworkCloudSim, GreenCloud, etc..) deal with the data as for size only without any consideration about the data allocation policy and locality. On the other hand, the NetworkCloudSim simulator is considered one of the most common used simulators because it includes different modules which support needed functions to a simulated cloud environment, and it could be extended to include new extra modules. According to work in this paper, the NetworkCloudSim simulator has been extended and modified to support data locality. The modified simulator is called LocalitySim. The accuracy of the proposed LocalitySim simulator has been proved by building a mathematical model. Also, the proposed simulator has been used to test the performance of the three-tire data center as a case study with considering the data locality feature.*

## KEYWORDS

*Cloud Computing, Data Locality, NetworkCloudSim Simulator*

## 1. INTRODUCTION

A Cloud is a type of distributed system consisting of a collection of interconnected and virtualized computers that are dynamically provisioned and presented as one or more unified computing resource(s). Because the cloud computing is considered a business model (i.e., it is based on pay-as-you-go principle), the provisioning of the resources depends on what is called Service-Level Agreements (SLAs) between the service provider and consumers [1, 2]. On the other hand, the cloud provider (CP) - person, or organization, or entity is responsible for providing available services to the interested parties, while the cloud broker (CB) manages the use, performance, and delivery of the cloud services. Also, he negotiates relationships between the cloud providers and the cloud consumers [3].

The cloud provides three types of service models; Software as a service (SaaS), platform as a service (PaaS), and infrastructure as a service (IaaS). The cloud deployment models are private, public, community and hybrid.

Now a day, large volumes of data are generated because of instrumented business processes, monitoring of user activity, website tracking, internet of things, accounting. Also, by progressing social network Web sites, the users create records of their lives by daily posting details of activities they perform. This intensive data is referred to Big Data. Big Data is characterized by what is referred to as a multi-V model; Variety, Velocity, Volume, and Veracity. Examples of Big Data include repositories with government statistics, historical weather information and forecasts, DNA sequencing, healthcare applications, product reviews and comments, pictures and videos posted on social network Web sites, and data collected by an Internet of Things [4].

MapReduce is a popular programming model for Big Data processing and analysis across the distributed environment using a large number of servers (nodes). The processing can occur on data, which are stored either in a filesystem (unstructured) or in a database system (structured). MapReduce supports data locality, where processing of data could be on or near the storage assets to reduce communication traffic. One of the important features of MapReduce is that it automatically handles node failures, hides the complexity of fault tolerance from the developers. MapReduce main functions are a map and reduce, where these functions are executed in parallel on the distributed environment [5, 6, 7, 8, 9, 10, 11]. On the other hand, MapReduce represents its power for processing large datasets with considering locality feature. Because MapReduce clusters have become popular these days, their scheduling is considered one of the important factors should be considered [12]. Hadoop is an open source implementation of Map Reduce. Hadoop as a Service is a cloud computing solution that makes medium and large-scale data processing accessible, easy, fast and inexpensive. This is achieved by eliminating the operational challenges of running Hadoop. Both Hadoop and Cloud have relation according to the Need.

Many open source cloud simulators like CloudSim, GreenCloud, NetworkCloudSim and CloudSimSDN have been introduced to implement and evaluate research approaches such as task scheduling, resource provisioning and allocation, security, and green cloud etc...CloudSimSDN simulator focuses on virtual machine provisioning according to the user defined software [13]. GreenCloud simulator deals with power consumption as the main factor [14]. Unfortunately, these simulators support specific research issues without any consideration about data locality. Therefore, NetworkCloudSim simulator provides different features which are needed for most research directions [15].

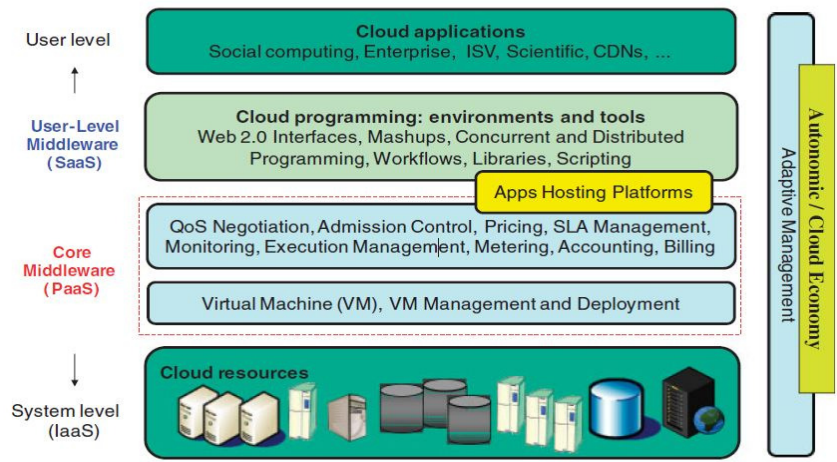

Figure 1. CloudSim architecture [16]

CloudSim simulator is the most used simulator because of its simplicity and flexibility. It is implemented using Java language without graphical user interface. It simulates the cloud activities at four layers that represent the services of the cloud. The first layer is user layer which supports activities of SaaS to the end user. So, the end user can configure the applications such as social network, research application, and other cloud applications. The second layer is the user-level middleware (SaaS) which supports user platform like the web interface, libraries, and workflow models. The third layer is the core middleware (PaaS) which supports access control, execution management, monitoring, and provisioning techniques, as well as, pricing. The fourth layer is the system level (IaaS) which supports the physical utilities of the cloud hardware such as powering, dynamic allocating, and resources distributions. CloudSim simulator includes modules for the most of the cloud's components such as virtual machine, data center, provisioning policy, and broker. Therefore, CloudSim becomes one of the most used simulators. Figure 1 shows the CloudSim architecture [16].

Data locality is concerned about where the data are stored in hosts contain storage devices. Data has two type of locality [17]:

- **Temporal Locality**; the last location of data accessed by the program,
- **Spatial Locality;** it is the permanent location of the data.

The placement techniques are used to distribute data across hosts based on availability, reliability, and Quality of Service (QoS) that the broker agrees to it with the users. Data locality affects the performance of any scheduling algorithm. If the scheduler fails to place the jobs near the data, extra time will be needed to transfer data depends on the network bandwidth. Therefore, the scheduling performance will be affected [18, 19]. On the other hand, data locality has three different locations; (1) the same host where no transfer time across the network is needed, (2) the same rack or switch; and (3) the remote host. In the case of the same rack and remote host, the job's time increases due data transfer across the network.

Unfortunately, the existed cloud simulators are not supported data locality. According to work in this paper, an extended Network Cloud Sim has been proposed to support data locality beside its functions. This extended simulator is called Locality Sim. According to the proposed Locality Sim, new resource management algorithms or models can be easily implemented, tested and evaluated.
The remainder of this paper is organized as follows: In Section 2, a survey of related work and briefly discussion about Network Cloud Sim and Cloud Sim SDN simulators are presented. In section 3, the architecture of the proposed Locality Sim simulator is introduced. In section 4, Locality Simassumptionsare discussed. The performance evaluation of the proposed Locality Sim simulator is discussed in section 5. Finally, the conclusion and future work are presented in 6.

## 2. RELATED WORK

Because NetworkCloudSim and CloudSimSDN based on CloudSim simulator, in addition, the proposed LocalitySim is an extension of NetworkCloudSim, NetworkCloudSim and CloudSimSDN will be discussed as a related work.

### 2.1. NetworkCloudSim Simulator

NetworkCloudSim is an extension to CloudSim simulator by adding some classes and extending other classes to enable the simulator to present real workload application, which consists of multi-task with each task consists of multi-stage [15]. NetworkCloudSim simulator provides scalable network and real workload application which improve the performance of the simulated data center. Figure 2 shows CloudSim architecture with NetworkCloudSim modification. According to NetworkCloudSim, each module is represented by class or more to act like real work and provides more control over each module. In addition, NetworkCloudSim represents the infrastructure of the data center by more than one component such as data center, host, switch, and storage. On the other hand, the components of the infrastructure have its related module and extensions to support provisioning, and scheduling policies. The main feature of NetworkCloudSim is the application module which supports real workload by dividing the application into a group of tasks with each task has a different type of states(i.e., send, receive, execute, and end). By using this application module, the most real applications become easy to be simulated.

## 2.2. CloudSimSDN Simulator

CloudSimSDN is another extension of the CloudSim simulator, but it focuses on virtual machine provisioning. CloudSimSDN simulator is used to evaluate the data center performance according to the user software-defined. CloudSimSDN provides a graphic user interface as one of input methods to configure the data center network.

Both NetworkCloudSim and CloudSimSDN simulators are considered popular because of their availability and holistic environment where many cloud components have been presented in modules and the interactions between them are managed. Unfortunately, both of them and other existed simulators are not supported data locality and even the effect of changing data location. Therefore, the simulated data center could not be able to measure the data allocation policy.

## 3. THE PROPOSED LOCALITYSIM ARCHITECTURE

Again here, the proposed LocalitySim simulator is an extension of CloudSim and a modified version of NetworkCloudSim with supporting data locality module. Figure 2 shows the architecture of the proposed LocalitySim simulator.

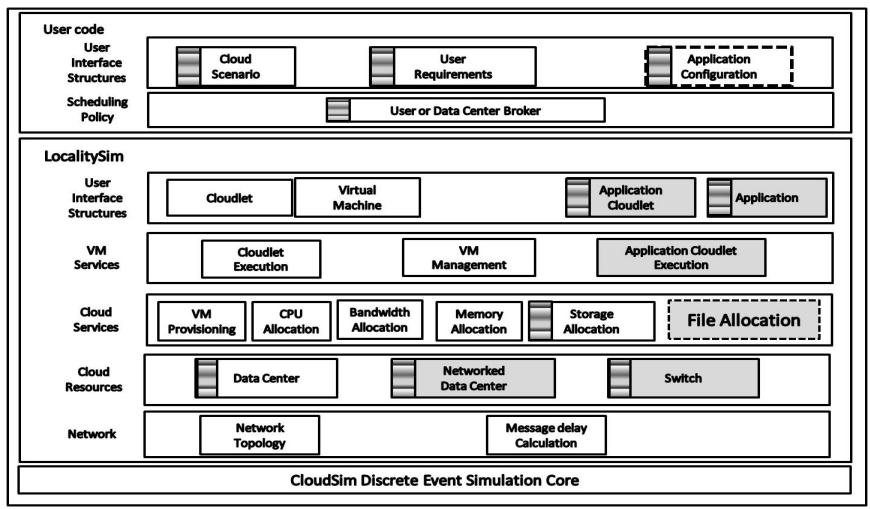

Figure 2. LocalitySim Architecture

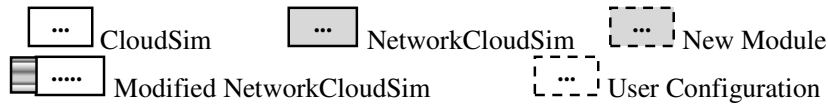

## 3.1. CloudSim Core Module

CloudSim core is used in the proposed LocalitySim simulator without any changes. It contains CloudSim Discrete Event Simulation Core with all modifications added by NetworkCloudSim. LocalitySim core layer contains the basic modules of the cloud simulator components such as a future queue, deferred queue, SimEntity, SimEvent and other basic modules. Future queue contains the jobs will be executed. When the job's time starts, it is transferred to Deferred queue. The components of the NetworkCloudSim are presented in Figure 3[15].

## 3.2. Data Center Module

IAAS is a cloud infrastructure as a service [3]. IAAS is the bottommost layer of the cloud services where the cloud resources exist. At this layer, the cloud presents allocation to cloud resources such as storage, network, and any computing resources as a pool of resources. Data center and network data center are modified to support the data locality. Providing hosts, virtual machine scheduler, bandwidth provisioning and RAM provisioning are implemented to create the new data center. According to work in this paper, Data center module is modified to support the data locality, by add name node module to data center object and networked data center extend the data center object with no change from NetworkCloudSim's network data center.

## 3.3 Switch Module

According to NetworkCloudSim, switch module simulates the function of the real switch. According to the switch module, the data delay on switches is calculated starting from the root switch which is considered the core of all switches at the networked data center. Only one root switch is considered to simplify the calculation and the network topology. The successor of the root switch is the aggregate switch with many child says edge switches. The aggregate switch acts as the main network data center clustering, while the edge has many child says hosts. According to work in this paper, the Switch module has been modified to support the data locality by determining the communication cost on the switches. This modification will be discussed in detail in section 5.

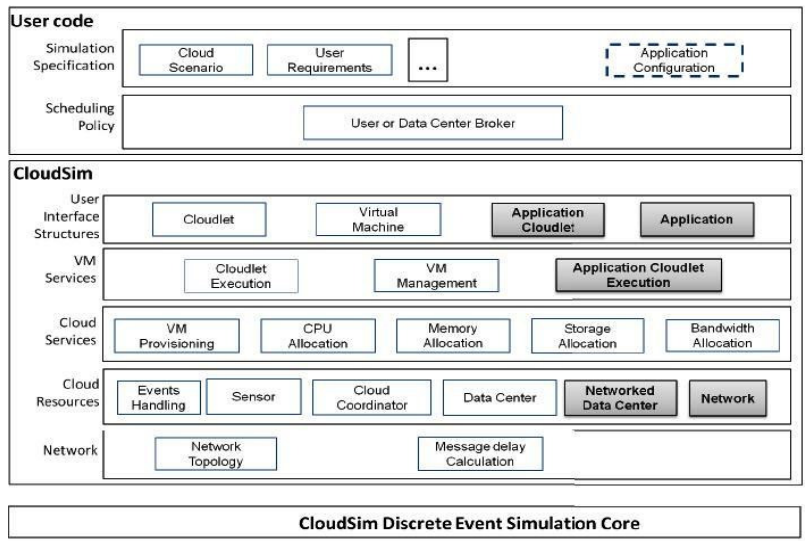

Figure 3. CloudSim architecture with NetworkCloudSim modification and extension [15]

## 3.4 Host Module

Host module simulates the work of real server or host machine which includes memory, storage, and processing elements. The host connects to other hosts across the three type of switches to group them in one pool of resources. Host module calculates the transfer cost or delays of moving data from the virtual machine to another in the same host. The communication cost of data depends on the location of the transferred data in case of a different host. The transferring data will be done in ascending order, where the data is moved from the host to another host on the same edge switch. Then, the data is transferred from host to another host in the same aggregate switch. Finally, data is transferred from one host to another host in the same root switch. The host contains San storage which contains the files belongs to the host. It has an provisioning policy for bandwidth and memory to allocate and divide the whole bandwidth and memory across host's virtual machine. It has virtual machine scheduling algorithm (e.g., time share – space share – and any customized algorithm) that responsible for allocating processing element to virtual machines [16]. The networked data center architecture is illustrated in Figure 4. According to the proposed LocalitySim simulator, the host module is modified to cover the data locality by calculating the inner communication cost on hosts (sender – receiver).

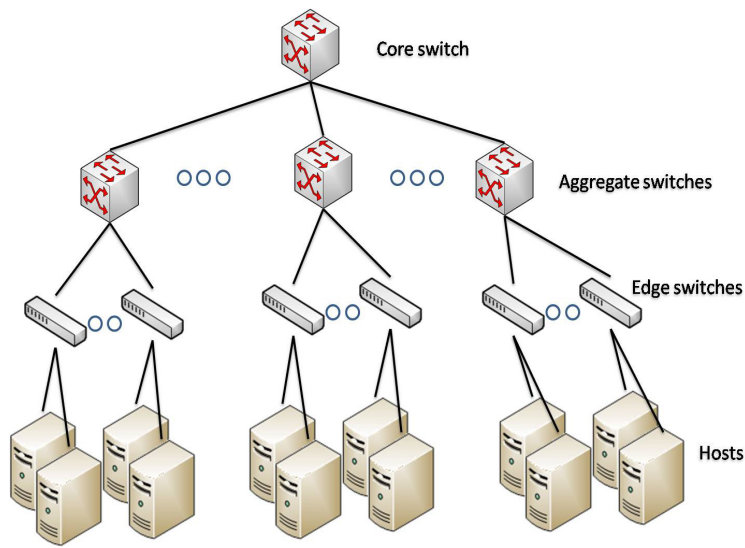

Figure 4. The networked data center architecture

### 3.5 Virtual Machine Module

The virtual machine (VM)is an abstraction of physical resources for executing the user's tasks[20]. The virtual machine module simulates the work of real VM. The main components of VM are memory and processing unit. The virtual machine module is responsible for provisioning VMs to hosts or schedule tasks on VMs. VM module contains the structure of VM, allocation policy, and scheduling algorithm.

### 3.6 File Allocation

File allocation module manages the file distribution on hosts and the search operation by using Name node implementation. Using data locality, the user can handle different type of file distributions, measure the impact of each file distributions, optimize data allocation policy, and accurate the performance measure of the real data center. File allocation module contains information about the location of each file, sender, receiver and the percentages of data locality types such as; 1) node locality, 2) rack (edge) locality,3) aggregate locality, and 4) root locality. Node locality means that the sender and receiver hosts are the same hosts (i.e., there is no data communication overhead between them). Rack (edge) locality means that data communication overhead will exist across the same edge switch that has the sender and receiver hosts. Aggregate locality means that data communication overhead exists across the same aggregate switch that has the sender and receiver hosts. Root locality means that data communication overhead exists across the same root switch that has the sender and receiver hosts. File allocation has been implemented by name node module to be the base of implementing data locality.

### 3.7 Application Cloudlet

Application cloudlet simulates the real application [15].It composes a group of network cloudlet that simulates the steps of the application or application's tasks. Each task or network cloudlet is composed of multistage at four states such as receive, send, execute and finish. by dividing the application into many parts, the user can simulate a lot of different applications which support the generality. Figure 5 shows the modelling of applications in the proposed LocalitySim simulator with data locality aware. There, Application cloudlet module has been modified to include data locality.

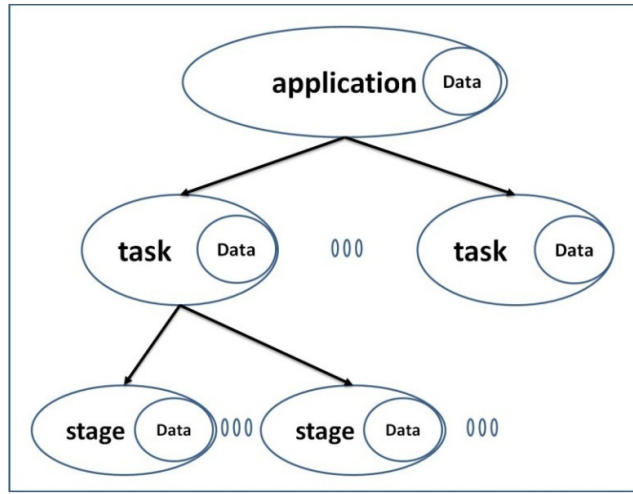

Figure 5. Modelling of Application cloudlet(locality wise)

### 3.8 Broker

The broker is an entity that manages the use, performance and delivery of cloud services and negotiates relationship between Cloud Providers and Cloud Consumers [3].Broker module simulates the work of cloud broker by calling appropriate modules and has all information about system components and requirements Broker is modified to manage the upgraded and new modules. According to the cloud scenario, Broker creates the virtual machines, distribute the data file across the hosts, generate the workload, and call the generated workload.

### 3.9 Cloud Scenario

Cloud scenario module illustrates the configuration of the cloud at IAAS and PAAS layers to initialize the simulator. The user can determine the number of hosts and VMS and their specifications. A graphic user interface (GUI) has been introduced as an input method to enter the user requirement parameters (number of jobs – data locality percentages – etc.) (See Figure 6).

### 3.10 Application Configuration

The application configuration is responsible for the used application structure in the simulator. Different types of applications can be implemented like multi-tier and message passing interface application. By extending or modifying the application cloudlet, the application configuration with considering data locality is done.

### 3.11 The User Requirements

The User requirements (i.e., RAM, the number of processing unit at each virtual machine – etc.) should be entered through LocalitySim GUI (see Figure 6). Cloud scenario, application configurations and user requirements are customized by the user. The customization could be using the GUI or by editing the source code.

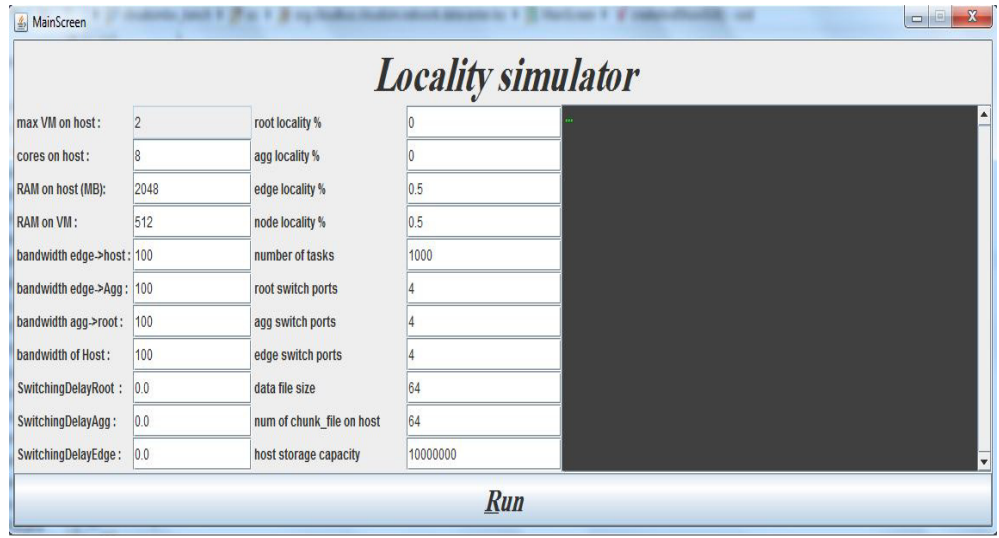

Figure 6. GUI of LocalitySim Simulator

## 3. LOCALITYSIM ASSUMPTIONS

There are some assumptions should be considered when using LocalitySim simulator tool. These assumptions are switches topology, and workload schema.

### 4.1 Switches Topology

LocalitySim has only one root switch with predefined ports. Also, it has a number of aggregate switches which are linked up to the root switch but not exceed the number of ports at root switch. Each aggregate switch is linked up to root switch and linked down to a number of edge switches. Edge switches depend on the number of ports on the aggregate switch. Edge switch is linked up to aggregate switch and linked down to many hosts depending on predefined ports at an edge switch (see Figure 4). Using the previous switches topology, the user can simulate data center with different topologies.

For simplicity, one copy of the chunk file will be considered at a data center

### 4.2 Workflow Schema

The default workflow application simulates the flow of application which consists of two tasks. The first task is used to execute and send data file. The second task is used to receive and execute data file. The two tasks simulate the process of reading the file from splitting files into the map functions. The workflow application is implemented at the class WorkflowApp, which can be modified or extended to change the structure of the required application. The file schema of a workflow application is a text file consisting of multi-lines, each line is an application, and contains three fields; application number, file number, and the identification of the virtual machine requests the file or the identification of virtual machine of map function.

## 5. LOCALITYSIM EVALUATION

The proposed LocalitySim simulates data center using three levels of switches; root switch, aggregate switches and edge switches. To prove the concept of the proposed LocalitySim simulator, an mathematical model of the data center has been built with considering a case study. The proposed mathematical model is a tree model with constraints as shown in Figure 7. The purpose of the mathematical model is to calculate the communication cost of data manipulation across the data center.

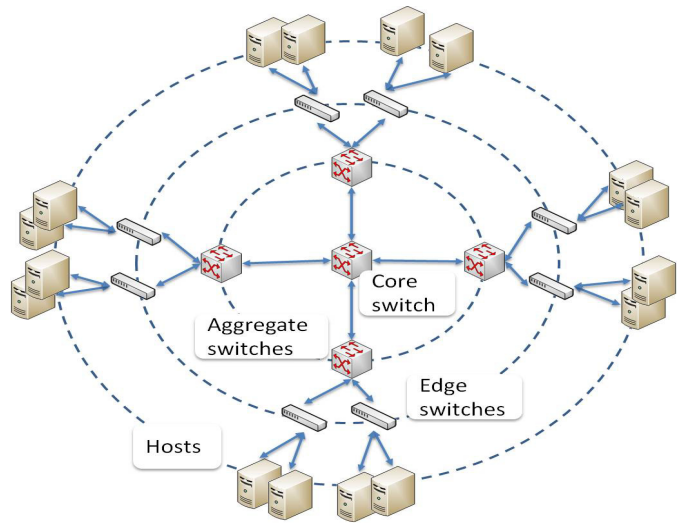

Figure 7. Mathematical model graph

## 5.1 LocalitySim Simulator Model

In this section, the principles of the LocalitySim tool are discussed

### 5.1.1 LocalitySim Graph (LSG)

LocalitySim Graph (LSG) is a tree graph of 12-tuple:
LSG = (N, NH, NSW, L,BW, D, C, FD,F,P, PATH,T)
Where:
1)   N = {n ∈ N : n >= 0 } - is a set of nodes
2)   NH= {nh∈ NH : n >= 0 }- is a set of hosts
3)   NSW= {nsw∈ NSW : n >= 0 }- is a set of switches
4)   L = { l$_{ij}$∈  L : i,j ∈N}- is a set of links between nodes
5)   BW = { bw$_{ij}$ ∈BW  : i,j∈ n , bw$_{ij}$>= 0 }- is a set of bandwidths
6)   D = { d$_{ij}$ ∈ D  :  i,j ∈ n , dij >= 0 }-  is a set of delays
7)   C= { c$_{ij}$ ∈ C  :  i,j ∈ n , dij >= 0 }- is the communication cost between nodes
8)   FD = {fd ∈ FD : fd >= 0  }- is a set of files
9)   F = { f$_{fd}$∈ F : f$_{fd}$> 0 , fd ∈FD}- is the file moves between node i and node j

10)     P = { $p_{ij} \in$ P : i,j $\in$ N ,fd $\in$ FD $p_{ij}$ is a 7-tuple, $p_{ij}$ = ($m_i,m_j,l_{ij},bw_{ij},d_{ij},f_{fd},c_{ij}$)  }- is a set of all moves at the datacentre and its communication cost

11)     PATH = {$path_{fd} \in$ PATH , $path_{fd}$ ={$p_{00},p_{01},p_{11}\ldots p_{(n-1)(n-1)},p_{(n-1)(n)},p_{(n)(n)}$} , n $\in$ N , fd $\in$ FD}- is a set of data paths

12)     T $=\sum_{i,j=0}^{n} c_{ij}(p_{ij})$ – the total   communication cost between nodes ( i, j) of existing path $P_{ij}$

The target is to calculate the communication cost of transferring the file size across the nodes.

### 5.1.2 Constraints and Mathematical Functions

$$T(LSG) = \sum_{i,j=0}^{n} c_{ij}(p_{ij}) \qquad (1)$$

$$c_{ij}(p_{ij}) = \begin{cases} \dfrac{f_{ij}}{bw_{ij}} + d_{ij}, & i \neq j \\[2mm] \dfrac{f_{ii}}{bw_{ii}} + d_{ii}, & i \in NH \\[2mm] d_{ii}, & i \in NSW \end{cases}$$

$$(2)$$

N = NH $\cup$ NSW          (3)
NH $\cap$ NSW = $\emptyset$        (4)

Equation (2) calculates one move of the file size. The move may be from node to another node or itself. The purpose of the mathematical model is to provide the effect of the data transferring between hosts at the datacentre. To express the movement form one host to another host, four cases are existed based on the locality types (i.e., node locality, rack (edge) locality, aggregate locality, and root locality)

1)    **node locality**;  the move from host to itself
| $path_{fd}$ | = |{$p_{ii}$} | = 1        (5)
2)    **rack locality**; the movement at the same rack switch
| $path_{fd}$ | = | { $p_{aa},p_{ab},p_{bb},p_{bc},p_{cc}$ } |=5     (6)
3)    **aggregate locality**; the movement at same aggregate switch
| $path_{fd}$ | = | { $p_{aa},p_{ab},p_{bb},p_{bc},p_{cc},p_{cd},p_{dd},p_{de},p_{ee}$ } |=9             (7)
4)    **root locality**; the movement at the same root switch
| $path_{fd}$ | = | { $p_{aa},p_{ab},p_{bb},p_{bc},p_{cc},p_{cd},p_{dd},p_{de},p_{ee},p_{ef},p_{ff},p_{fg},p_{gg}$} |=13       (8)

### 5.1.3 Data Locality Proof

$c_{aa}(p_{aa}) = c_{nn}(p_{nn})$   a,n$\in$ NH         (9)
$c_{bb}(p_{bb}) = c_{dd}(p_{dd})$  b,d $\in$ NSW     (10)
$c_{ab}(p_{ab}) = c_{de}(p_{de})$  a,b,d,e $\in$ N     (11)
$c_{ab}(p_{ab}) = c_{ba}(p_{ba})$  a,b $\in$ N          (12)
$\sum_{i=0}^{n} C_{ii}(p_{ii})$ = H where i $\in$ NH      (13)
$\sum_{i=0}^{n} C_{ii}(p_{ii})$ = SW where i $\in$ NSW   (14)
$\sum_{i,j=0}^{n} C_{ij}(p_{ij})$ = CH where i,j $\in$ N     (15)

$\forall$**Equations from 1 to 15**$\therefore$

$$T(LSG) = \begin{cases} H, & All\ node \\ 2H + SW + 2CH, & All\ rack \\ 2H + 3SW + 4CH, & All\ aggregate \\ 2H + 5SW + 6CH\ , & All\ root \end{cases}$$

$$(16)$$

if $\forall\ d_{ij} \in D$ , $d_{ij} = 0$

Moreover, $\forall\ bw_{ij} \in BW, bw_{kl} \in BW, bw_{ij} = bw_{kl}$

*then*, Equation (16) will be as especial case:

$$T(LSG) = \begin{cases} H, & node \\ 4H, & rack \\ 6H, & aggregate \\ 8H, & root \end{cases}$$

$$(17)$$

The importance of data locality is defined by this constructive proof, where the communication cost of data manipulating is defined using equations (16), (17).

According to equation (17), Figures 8 represents the mathematical model communication cost percentage.

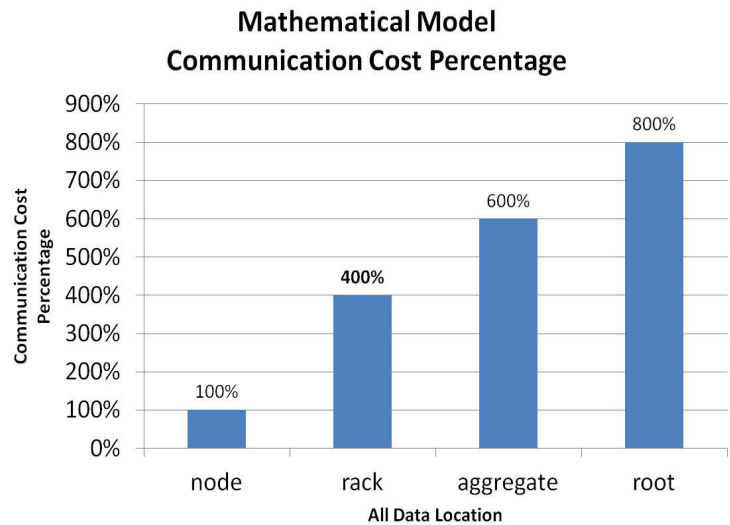

Figure 8. Mathematical model communication cost percentage

## 5.2. Case Study

In this case study, LocalitySim simulator simulates only the map function of the map-reduce programming model, which reads the data from storage across the data center.

By considering different values of the proposed LocalitySim parameters (bandwidth, the number of tasks, the number of aggregate switches, the number of edge switches, the number of hosts), the communication cost will be determined.

**Experiment One**

Assuming the parameters' values are represented in Table 1.

Table 1. Assumption of LocalitySim's parameters

| Item | Value |
|------|-------|
| All bandwidth of any node to another | equal |
| All bandwidth | 100 MB |
| Delay | 0 |
| Number of tasks | 1000 |
| Chunk file size | 64 MB |
| Number of Switch root | 1 |
| Number of Aggregate switches | 4 |
| Number of Edge switch | 16 |
| Number of hosts | 64 |

The communication cost for each locality type(i.e., node locality, rack locality, aggregate locality, and root locality)is represented in Figure 9.

By comparing the results of the mathematical model and the case study results, it is found that the case study results agree with the mathematical model (see Figures 8, 9).

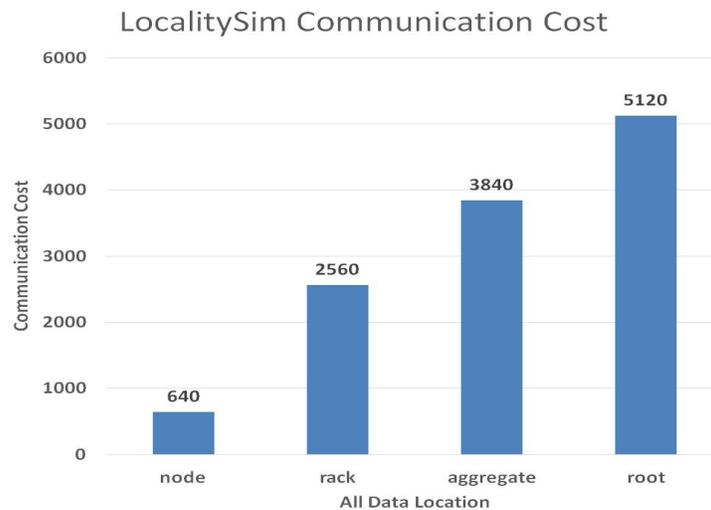

Figure 9. Result of Experiment one

**Experiment Two**

Assuming the parameters' values are represented in Table 2. The communication cost for each locality type is represented in Figure 10. Again here, the results of the mathematical model and the case study results are agreed (see Figures8, 10).

Table 2. Assumption of LocalitySim's parameters

| Item | Value |
|------|-------|
| All bandwidth of any node to another | equal |
| All bandwidth | 1000 MB |
| Delay | 0 |
| Number of tasks | 2000 |
| Chunk file size | 64 MB |
| Number of Switch root | 1 |
| Number of Aggregate switches | 6 |
| Number of Edge switch | 24 |
| Number of hosts | 96 |

Therefore, the experimental results of the case study using different values of the proposed LocalitySim parameters (bandwidth, the number of tasks, the number of aggregate switches, the number of edge switches, and the number of hosts) are agreed with the mathematical model results.

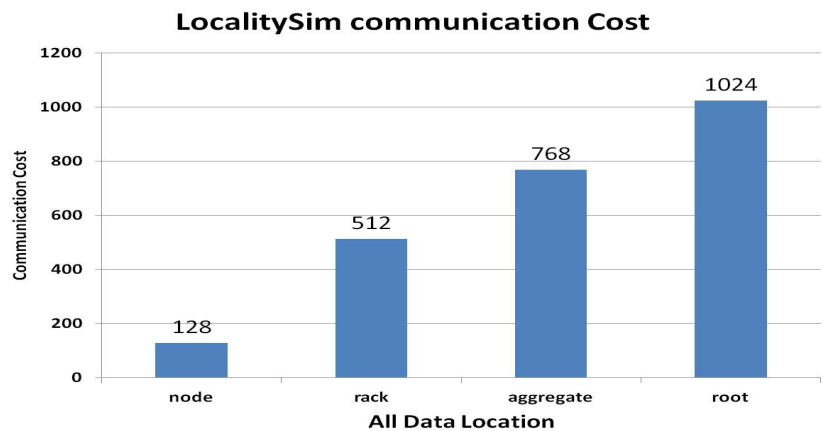

Figure 10. Results of experiment two

Table 3 illustrates the features of the proposed LocalitySim tool with respect to GreenCloud tool, and NetworkCloudSim tool. According to Table 3, it is noticed that the proposed LocalitySim tool demonstrates the importance of the data locality at the datacenter's efficiency.

Table 3. Simulator Comparison

| Item | NetworkCloudSim | CloudSimSDN | LocalitySim |
|------|-----------------|-------------|-------------|
| language | Java | java | Java |
| availability | Open source | Open source | Open source |
| GUI | no | yes | yes |
| Communication models | full | full | full |
| Data locality | no | no | yes |
| Data Centers | single | multi | single |

## 6. CONCLUSIONS

The existed open source cloud simulators like CloudSim, GreenCloud, NetworkCloudSim and CloudSimSDN are not considered data locality. According to work in this paper, the LocalitySim simulator has been introduced with considering the data locality. Therefore, the effect of the data locality types, distributing the file across the hosts and the topology of the data center can be simulated.

As a future work, the effect of data locality type, application structure, and the network topology could be studying at the same time to investigate the effect of data locality in the efficiency of the datacenter.